\documentclass{article}
\usepackage[T1]{fontenc}
\usepackage[latin9]{inputenc}
\usepackage{amssymb}
\usepackage{graphicx}

\makeatletter

\newcommand{\noun}[1]{\textsc{#1}}

\usepackage{amsfonts}

\providecommand{\U}[1]{\protect\rule{.1in}{.1in}}
\providecommand{\U}[1]{\protect\rule{.1in}{.1in}}
\providecommand{\U}[1]{\protect\rule{.1in}{.1in}}
\providecommand{\U}[1]{\protect\rule{.1in}{.1in}}
\providecommand{\U}[1]{\protect\rule{.1in}{.1in}}
\providecommand{\U}[1]{\protect\rule{.1in}{.1in}}
\providecommand{\U}[1]{\protect\rule{.1in}{.1in}}
\providecommand{\U}[1]{\protect\rule{.1in}{.1in}}
\providecommand{\U}[1]{\protect\rule{.1in}{.1in}}
\providecommand{\U}[1]{\protect\rule{.1in}{.1in}}
\providecommand{\U}[1]{\protect\rule{.1in}{.1in}}
\providecommand{\U}[1]{\protect\rule{.1in}{.1in}}

\makeatother

\begin{document}

\title{Interference Effects in Potential-Wells }

\author{W. J. Mullin $^{a}$ and F. Lalo\"{e} $^{a}$}
\maketitle
\begin{abstract}
We propose using an array of potential wells as an interferometer,
in which the beam splitters are provided by tunneling during an appropriate
time through the barrier between wells. This arrangement allows demonstration
of generalized Hong-Ou-Mandel effects with multiple particles traversing
one or several beam splitters. Other interferometer effects can occur, including
a violation of the Bell-Clauser-Horne-Shimony-Holt form of the Bell
inequality. With interactions, one sees various effects including so-called
fermionization, collective tunneling, and self-trapping. 
\end{abstract}
$^{a}$Department of Physics, University of Massachusetts, Amherst,
Massachusetts 01003 USA\\
$^{b}$ Laboratoire Kastler Brossel, ENS, UPMC, CNRS ; 24 rue Lhomond,
75005 Paris, France

\smallskip{}

\section{Introduction}

The Hong-Ou-Mandel (HOM) effect \cite{HOM,Ou} is perhaps the simplest
example of boson-boson interference, in which two particles approaching
in two different inputs of a beam splitter always exit both in the
same output, because destructive interference removes the possibility
of exits in separate outputs. Experimental verifications have
been carried out with photons \cite{HOM,HOMOther}. Other cases with
two photons from each side \cite{Ou2} as well as two from one side
and one from the other have also been considered theoretically \cite{Wang},
and experimentally seen \cite{Sanaka}. Theoretical generalizations using cold
atoms have recently been considered, including one in which Fock states
of arbitrary numbers of particles impinge on a beam splitter \cite{LM,ML},
and another using pair-correlated atoms produced via a collision of
two Bose-Einstein condensates \cite{Kher}. 
Recent experiments using particles rather than photons have
involved electrons \cite{elect1}, $^{87}$Rb atoms trapped in optical
tweezers \cite{Kauf}, and helium atoms \cite{Westb} where the beam
splitter was an Bragg scattering optical grating. Each of the papers
using matter waves rather than photons emphasizes that experiments
developing coherent indistinguishable pairs of particles may be relevant
in other areas such as quantum computing and information processing
\cite{compute}, highly sensitive force detection \cite{Force}, quantum
simulations \cite{Simu}, testing Bell inequalities with material observables
\cite{Kher2}, etc . Here we want to consider HOM-like interference
with the use of Fock states of cold atoms undergoing simple tunneling
in optical double (or multiple) potential-wells in a method analogous
to that experimentally used in Ref. \cite{Kauf}. 

The essential element in the HOM effect, or in any interferometer,
is a beam splitter. In the case of cold gases, various forms of beam
splitter have been designed, including, e.g., Bragg scatterers \cite{Westb, Bragg}
and double potential-well devices. In the latter case, the usual format
\cite{DoubWell} involves guiding a matter wave through a potential
well having a spatially or temporally growing central repulsive peak
that divides the beam into two parts, which, for example, can be recombined
further along to show interference effects analogous to a two-slit
interferometer. In one case the authors \cite{Andersson} proposed
using a pair of side-by-side wave guides having a narrow region where
tunneling took place; by adjusting the wave packet width (or momentum)
and the barrier width the packet splits into two equal parts, thus
using the time of tunneling as the main element to form a 50-50 beam
splitter. Recently Daley et al \cite{Daley} suggested using double-well
tunneling as a beam splitter to measure the order-2 R\'{e}nyi entropy
of an entangled state. Compagno et al \cite{Campagno} considered
boson and fermion walkers on a one-dimensional lattice with an optical
impurity with transmission and reflection coefficients adjusted to
act as a beam splitter. Here we will further examine the
possible use of the time of tunneling as a beam splitter, a method
already adopted by Ref. \cite{Kauf} in their experiment.

Most interference experiments with cold atoms have involved observing
a periodicity in the particle density as a function of position or
angle. An alternative would be counting the number of particles in
the detectors placed after a beam splitter. In the case of the photon HOM
experiment a minimum in the coincidence between the two detectors
was observed. However, there has been considerable recent progress
in the actual observation of individual atoms via the ``quantum gas
microscope'' \cite{Greiner,Bloch}. In this case the ``approach
is to assemble quantum information systems with full control over
all degrees of freedom, atom by atom, ion by ion'' \cite{Greiner}.
We will assume such an approach is literally possible so that the
number of atoms in each well is known at the beginning of the experiment
(input to the beam splitter) and at the end in the detectors after
the beam splitters. As in Ref. \cite{Kauf} we assume that
the beam splitter involves turning on tunneling between two or more
wells for a set time. In this experiment  two atoms were trapped in their
ground states in wells with optical tweezers and observed  via
fluorescence. The distance and barrier between the wells could be
varied to enhance or cut-off tunneling. In Ref \cite{Westb} Bragg
scattering was used for mirrors and a beam splitter with observation
of pairs or singles of atoms  falling on a multipixel detector. We
will look at variations of the HOM effect but also see that these
remarkable experimental methods may yield other interesting possibilities
where more elaborate interferometry is possible.

\section{Double-well beam splitter}

\begin{figure}[h]
\centering \includegraphics[width=3.5in]{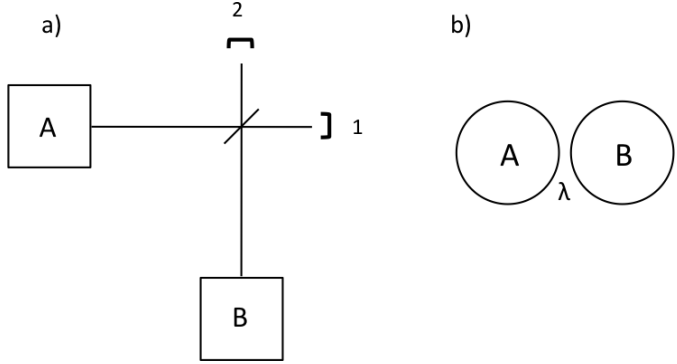}

\caption{a) $N_{A},N_{B}$ bosons proceed from the sources A and B to a beam
splitter, followed by two detectors 1 and 2, where $m_{1}$ and $m_{2}$
particles are detected. b) A double-well potential with tunneling
strength $\lambda$ between cells A and B. }

\label{fig:TwoTypes}
\end{figure}

Fig. \ref{fig:TwoTypes}(a) shows a beam splitter with two particle
sources and two detectors. In the standard beam splitter with the
annihilation operator for source $A$ being $\alpha$ and that for
$B$ is $\beta$ then the annihilation operators for detectors 1 and
2 are, respectively
\begin{eqnarray}
d_{1} & = & \frac{1}{\sqrt{2}}(\alpha+i\beta)\nonumber \\
d_{2} & = & \frac{1}{\sqrt{2}}(i\alpha+\beta)
\end{eqnarray}

We want to show that the the double well of Fig. \ref{fig:TwoTypes}(b)
is equivalent to the device of \ref{fig:TwoTypes}(a) if the tunneling
occurs for a set time period. Let $a$ be the destruction operator
for a particle in well A and $b$ for one in well B. Then the two-level
Hamiltonian we consider is taken as
\begin{equation}
H/\hbar=E_{0}(a^{\dagger}a+b^{\dagger}b)-\lambda(a^{\dagger}b+b^{\dagger}a)+\frac{W}{2}(a^{\dagger}a^{\dagger}aa+b^{\dagger}b^{\dagger}bb)\label{eq:H}
\end{equation}
where we have assumed the wells are at equal depth with identical
particles in each. The tunneling parameter is $\lambda$ with interactions
of strength $W$ among particles in the same well. In the double-well
case with no interactions the annhilation operators evolve in time
according to 
\begin{eqnarray}
a(t) & = & a(0)\cos\lambda t+ib(0)\sin\lambda t\nonumber \\
b(t) & = & ia(0)\sin\lambda t+b(0)\cos\lambda t
\end{eqnarray}
At time $t=\pi/4\lambda$ we have 
\begin{eqnarray}
a\left(\frac{\pi}{4\lambda}\right) & = & \frac{1}{\sqrt{2}}\left(a(0)+ib(0)\right)=d_{1}\nonumber \\
b\left(\frac{\pi}{4\lambda}\right) & = & \frac{1}{\sqrt{2}}\left(ia(0)+b(0)\right)=d_{2}
\end{eqnarray}
which is then equivalent to the 50-50 beam splitter. The simple two-particle
HOM effect, with a single particle initially in each well, corresponds
to the following state vector:
\begin{equation}
a^{\dagger}\left(\frac{\pi}{4\lambda}\right)b^{\dagger}\left(\frac{\pi}{4\lambda}\right)\left|0\right\rangle =\frac{1}{2}\left(a(0)^{\dagger}-ib(0)^{\dagger}\right)\left(-ia(0)^{\dagger}+b(0)^{\dagger}\right)\left|0\right\rangle =\frac{-i}{2}\left(a(0)^{\dagger}a(0)^{\dagger}+b(0)^{\dagger}b(0)^{\dagger}\right)\left|0\right\rangle 
\end{equation}
that is, ending up with a superposition of a pair of particles in
each well as expected in the simplest HOM effect. The HOM effect arises
because of the destructive interference resulting from exchange of
particles in the final state. 

For the general case of arbitrary $N$ we can solve for the coefficients
in the expansion of the wave function $\left|\psi(t)\right\rangle $
of $H$ in Fock states $\left|n,N-n\right\rangle $, having $n$ particles
in well A and $N-n$ in well B. If we take
\begin{equation}
\left|\psi(t)\right\rangle =\sum_{n}c_{n}(t)\left|n,N-n\right\rangle 
\end{equation}
then the differential equation for $c_{n}(t)$ can be shown to be
\begin{eqnarray}
i\frac{dc_{n}(t)}{dt} & = & \sum_{n^{\prime}}\left\langle n,N-n\right|H/\hbar\left|n^{\prime},N-n^{\prime}\right\rangle c_{n^{\prime}}(t)\nonumber \\
 & = & -\lambda\left[\sqrt{n(N-n+1)}c_{n-1}(t)+\sqrt{(n+1)(N-n)}c_{n+1}(t)\right]\nonumber \\
 &  & +\frac{W}{2}\left[n^{2}-n+(N-n)^{2}-(N-n)\right]c_{n}(t)\label{eq:cnEqs}
\end{eqnarray}
where we have omitted the term in $E_{0}$ that leads to a trivial
phase factor.

\section{HOM calculations\label{sec:HOM-calculations}}

\begin{figure}[h]
\centering \includegraphics[width=3in]{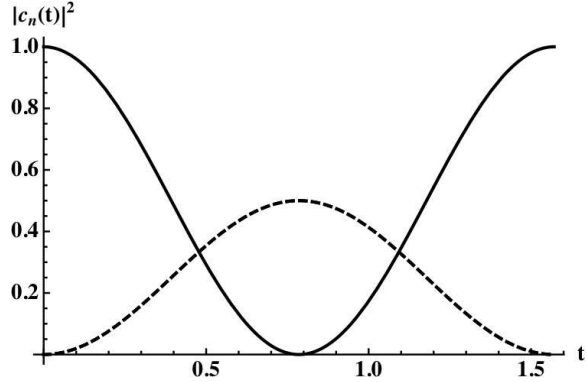}

\caption{Double-well HOM result with $N=2$, $N_{A}=N_{B}=1,$ showing the
time dependence of the probabilities of particle occupation of the
wells with no interaction, $\gamma=0$. The probabilities are shown
with $|c_{1}(t)|^{2}$ as a solid line, and $|c_{0}(t)|^{2}=|c_{2}(t)|^{2}$,
dashed. Initially we assume the wells each contain one particle: $c_{1}(0)=1,$
and $c_{0}(0)=c_{2}(0)=0$. After time $t=\pi/4$ (in units of $\hbar/\lambda$)
the probabilities become $|c_{1}|^{2}=0,$ and $|c_{0}|^{2}=|c_{2}|^{2}=1/2.$ }

\label{fig:HOM11Time}
\end{figure}

We can use Eqs. (\ref{eq:cnEqs}) to compute HOM-type interferences.
For the simplest case of $N_{A}=N_{B}=1,$ the equations are
\begin{eqnarray}
i\dot{c}_{0} & = & \gamma c_{0}-\sqrt{2}c_{1}\nonumber \\
i\dot{c}_{1} & = & -\sqrt{2}c_{0}-\sqrt{2}c_{2}\label{eq:Neq2}\\
i\dot{c}_{2} & = & \gamma c_{2}-\sqrt{2}c_{1}
\end{eqnarray}
where now we have expressed the time in units of $\hbar/\lambda$
and defined $\gamma=W/\lambda.$ The solutions of these equations
for the probabilities $|c_{n}(t)|^{2}$ are shown in Fig. \ref{fig:HOM11Time}

For $\gamma=0$ the analytic solutions are
\begin{eqnarray}
c_{0}(t) & = & c_{2}(t)=\frac{i}{\sqrt{2}}\sin2t\nonumber \\
c_{1}(t) & = & \cos2t
\end{eqnarray}
giving the wave function at time $t=\pi/4$ as
\begin{equation}
\left|\psi(t=\pi/4)\right\rangle =\frac{i}{\sqrt{2}}\left(\left|0,2\right\rangle +\left|2,0\right\rangle \right)
\end{equation}
 Thus the probabilities for the HOM effect, with 
\begin{equation}
p_{n}\equiv|c_{n}(\pi/4)|^{2}
\end{equation}
are the standard results: $p_{0}=p_{2}=1/2;$ $p_{1}=0.$

\begin{figure}[h]
\includegraphics[width=3in]{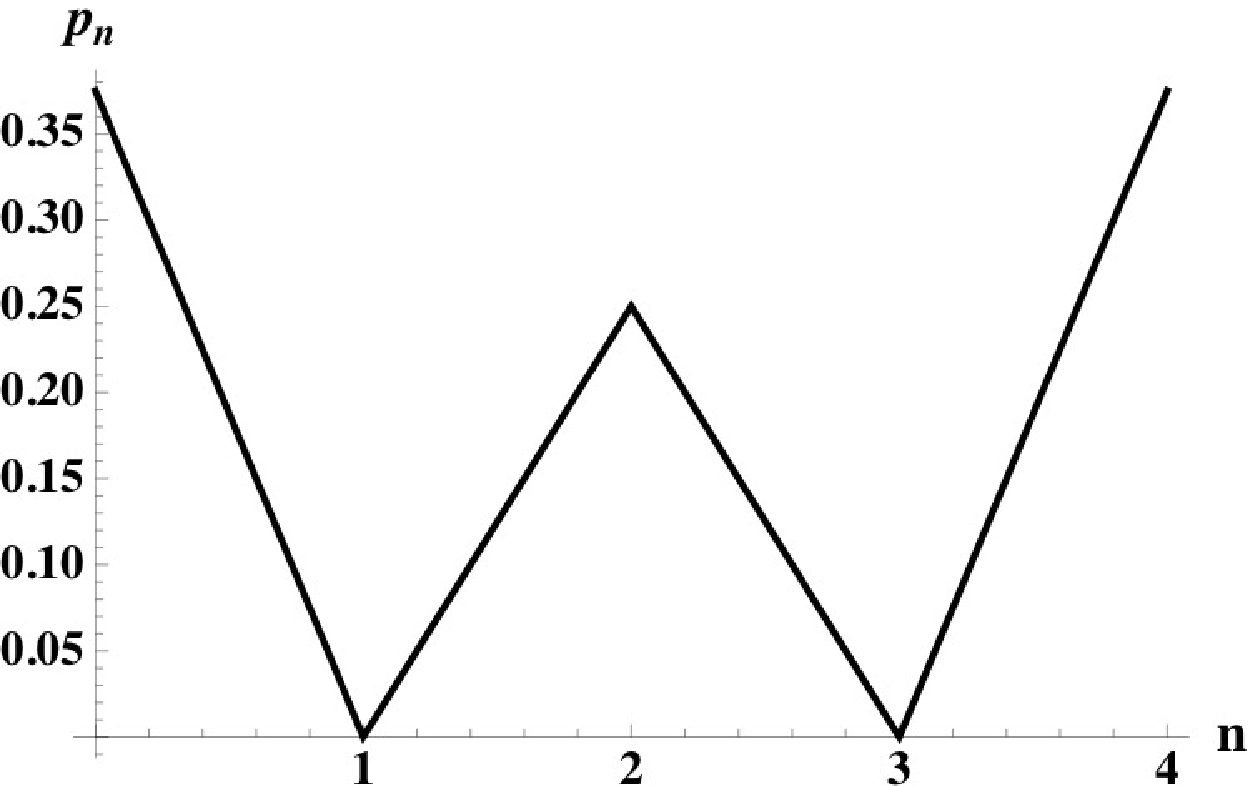}$\quad\quad$\includegraphics[width=3in]{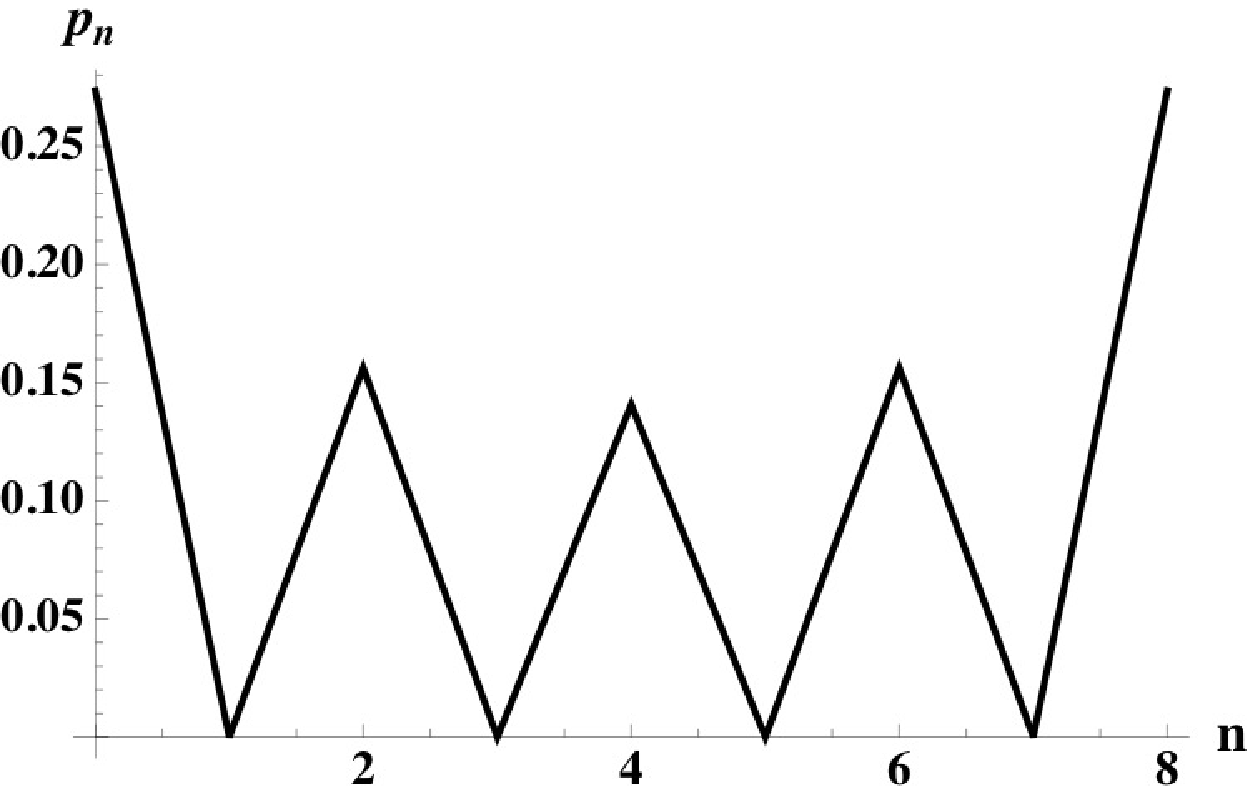}

\caption{Double-well HOM result with $N=4$ and 8. (Left) Probabilities $p_{n}$
of particle occupation of the well A at time time $t=\pi/4$, for
initial conditions $N_{A}=N_{B}=2$; the interaction parameter $\gamma=0$.
(Right) Similarly for initial conditions $N_{A}=N_{B}=4.$}

\label{fig:HOM422And44}
\end{figure}

\begin{figure}[h]
\centering \includegraphics[width=3in]{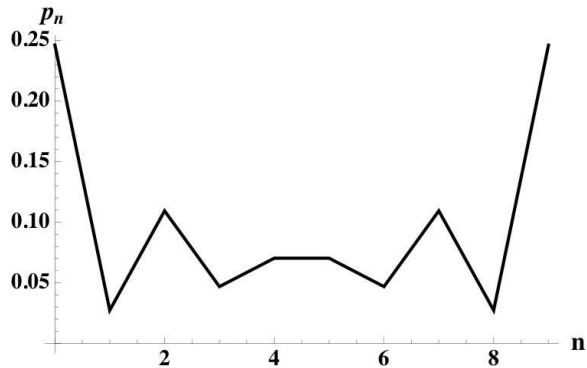}

\caption{Double-well HOM result for $N=9$, with $N_{A}=4$, $N_{B}=5$ initially,
showing the probabilities $p_{n}$ of particle occupation of the well
A at time time $t=\pi/4$. }

\label{fig:HOM45}
\end{figure}

This ``HOM time'' is just as identified in the experiment
of Ref. \cite{Kauf}, but in our case, the HOM effect can be generalized to more particles in each source
as was done in Refs. \cite{LM,ML}. In Fig. \ref{fig:HOM422And44}
we show the cases of $N=4$ and 8, where again we consider the occupations
of the wells at time $t=\pi/4$ with equal occupation in each well
initially. 
If the initial number of particles is not the same in each well then
the HOM preference for even occupation can't be maintained so purely
as seen in Fig. \ref{fig:HOM45}.

\begin{figure}[h]
\centering\includegraphics[width=3in]{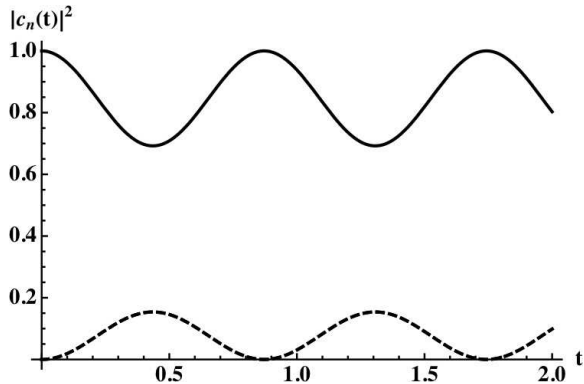}

\caption{Double-well HOM probabilities $|c_{n}(t)|^{2}$ as a function of time
for $N=2$ (initially $N_{A}=N_{B}=1$) with $\gamma=6$; $|c_{0}(t)|^{2}$,
$|c_{2}(t)|^{2}$ shown as dashed lines; $|c_{1}(t)|^{2}$ as solid
line. Compare with Fig. \ref{fig:HOM11Time}, for which $\gamma=0.$
As $\gamma$ is increased beyond the value here, the amplitudes of
the oscillations diminish further. }

\label{fig:Fermionization}
\end{figure}

Until now we have reproduced the results of Ref. \cite{LM} exactly.
One aspect of the present approach to the HOM analysis is that we
can include the effects of interactions. In the $N=2$ case the probabilities,
gotten from a numerical solution of Eqs. (\ref{eq:Neq2}), show that
$p_{0}$ and $p_{2}$ decrease as $\gamma$ increases and $p_{1}$
increases until all probabilities are equal at $\gamma\approx2.5$.
By $\gamma=6$, $p_{0}$ and $p_{2}$ are very small. This might be
expected for a repulsive potential, but the result, (and the Eqs.
(\ref{eq:cnEqs}) for any time) are invariant under sign reversal
of $\gamma$. The results are the same as if the particles had been
fermions. Indeed such so-called\emph{\noun{ }}\emph{fermionization}
of the system has been seen previously \cite{Campagno,fermization}.
We can see what is happening by looking at the full time dependence
in Fig. \ref{fig:Fermionization}. 
The average probability for staying in the original \{1,1\} state
is enhanced and those for the pair states \{2,0\} and \{0,2\} are
diminished. This result illustrates coordinated tunneling as seen
in other work \cite{fermization,PairTunnelA,PairTunnelB}. In Refs.
\cite{fermization} and \cite{PairTunnelB} two particles were started
out in the same well, and the result was the pair states became favored
and the \{1,1\} state diminished in probability. In our case any coordinated
tunneling would seem to involve the pairs crossing together, but in
opposite directions, so that we get the opposite result. In Ref. \cite{Campagno}
bunching was seen for two non-interacting bosons and antibunching
for a pair of fermions or strongly repulsive bosons.

\begin{figure}[h]
\includegraphics[width=2in]{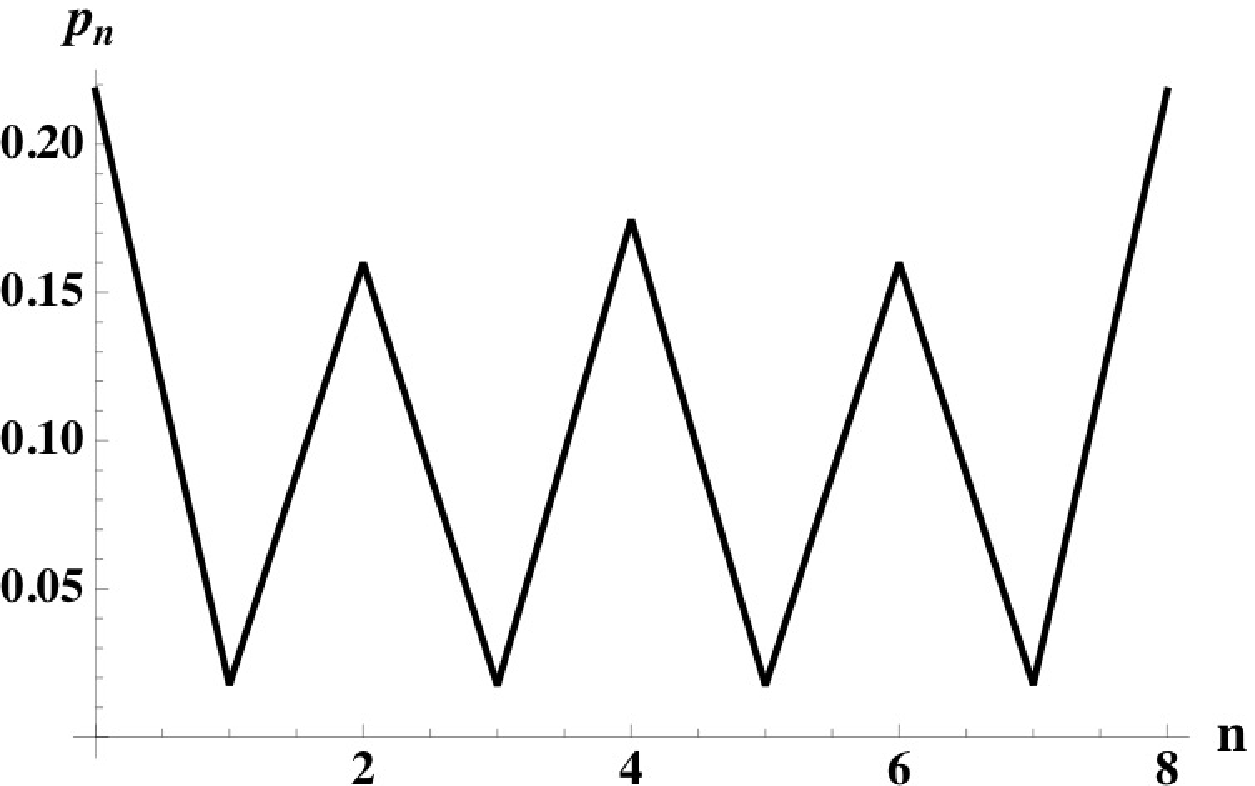}$\quad\quad$\includegraphics[width=2in]{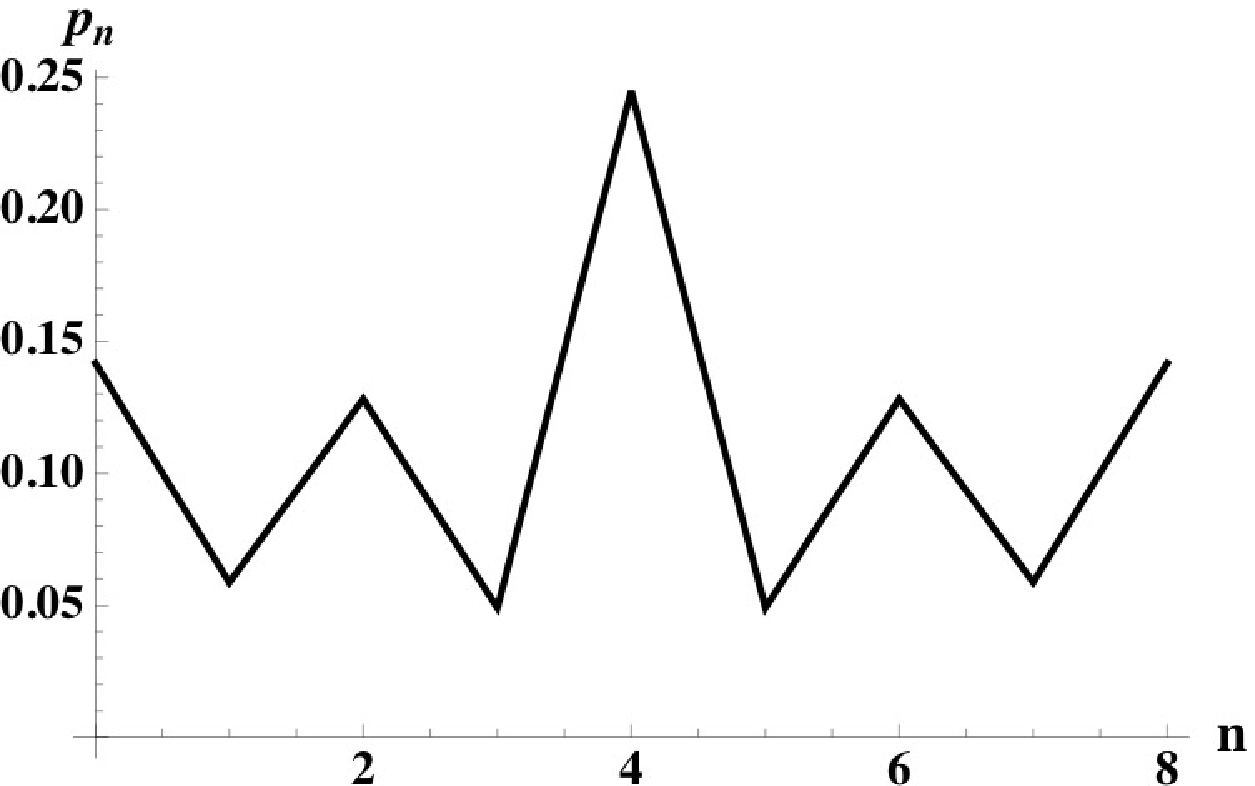}$\quad\quad$\includegraphics[width=2in]{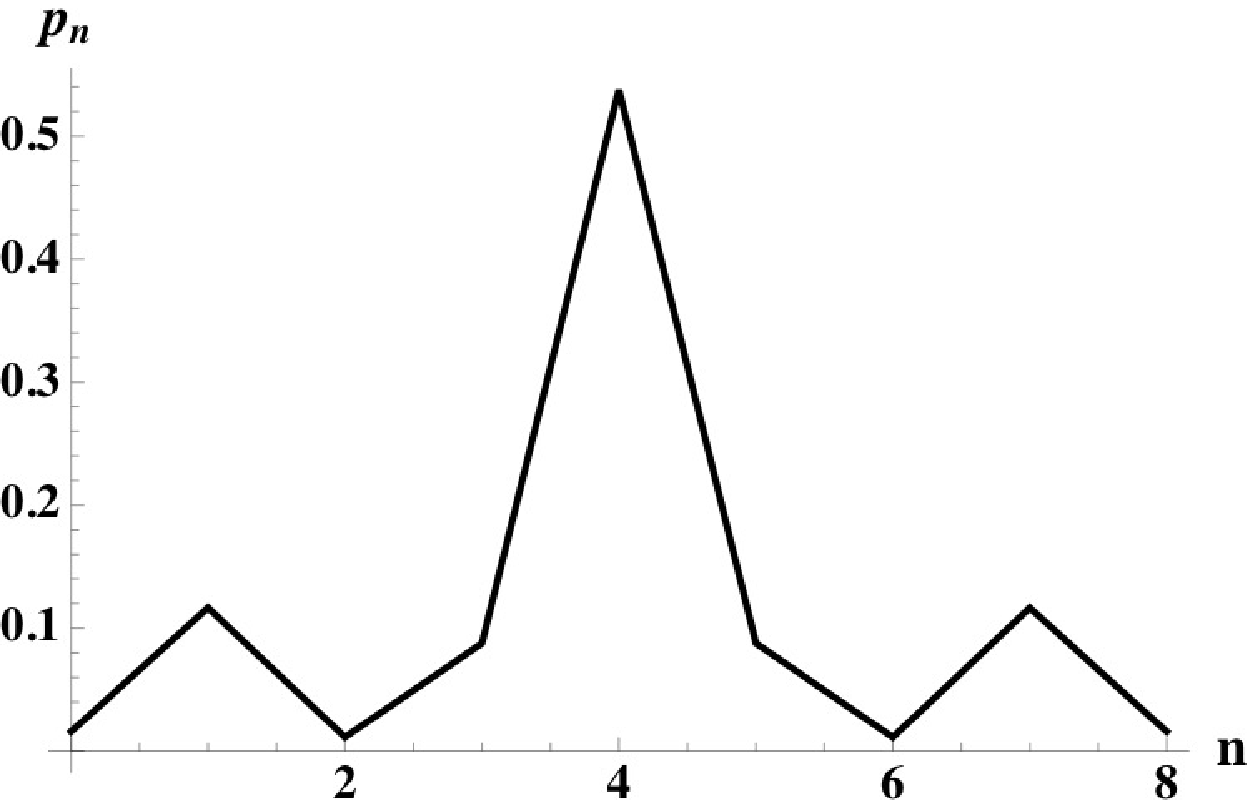}

\caption{Effects of interactions $\gamma$ on the double-well HOM results with
$N=8$. (Cf. Fig. \ref{fig:HOM422And44}(Right)). (Left) Probabilities
$p_{n}$ of particle occupation of the well A for initial conditions
$N_{A}=N_{B}=4$, and interaction parameter $\gamma=0.3$. (Center)
Similarly with $\gamma=0.5$. (Right) Similarly for $\gamma=1.$}

\label{fig:GammaN8}
\end{figure}

In Fig. \ref{fig:GammaN8} we show the effect of increasing $\gamma=W/\lambda$
on the initially equal-sided $N=8$ case. At small $\gamma=0.3$ the
HOM effect becomes less ``pure'' with odd states \{1,7\},\{3,5\},
etc now possible; by $\gamma=0.5$ the \{4,4\} state is enhanced;
and at large $\gamma=1$ the \{4,4\} state has become highly favored.
The effect seems related to ``self-trapping'' in which the sign
of an initial imbalance in the number of particles in the two wells
becomes fixed when the interaction parameter exceeds a critical value
\cite{PairTunnelB,SelfTrapA,SelfTrapB,SelfTrapC,SelfTrapD}, with
reduced amplitude oscillations away from the initial value of the
imbalance. We discuss some results of an analysis of self-trapping
in the Appendix and how our results might be similar. For fixed $N=8$
the critical interaction strength for self-trapping is $\gamma=0.5$;
the graphs then show results below, at, and above this critical value.
This effect is again independent of the sign of $\gamma.$ 

\section{Generalizations}

\begin{figure}[h]
\centering \includegraphics[width=3.5in]{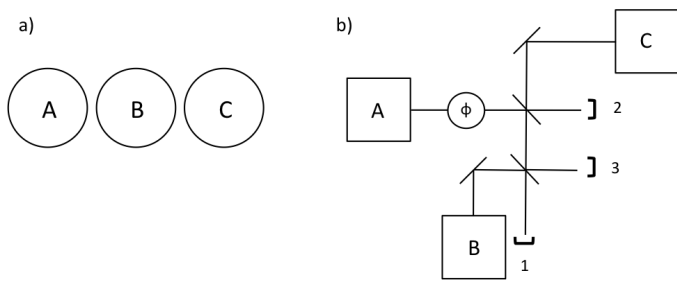}

\caption{(a) Three wells in a line. (b) The interferometer equivalent to the
three-well line of (a). The phase shift needed is $\phi=\pi.$ }

\label{figThreeWells}
\end{figure}

In addition to simulating a single beam splitter, particle tunneling
between wells can behave as an interferometer. Ref. \cite{Andersson2}
coupled two wave guides together with tunneling at two points, corresponding
to beam splitters, to make a two-input, two-output interferometer.
Suppose we put three wells in a line as shown in Fig. \ref{figThreeWells}a.
Particles in the outer wells A and C are connected by tunneling to
the center well B. The tunneling Hamiltonian, without interactions,
is
\begin{equation}
H/\hbar=\lambda(a^{\dagger}b+ab^{\dagger}+b^{\dagger}c+bc^{\dagger})
\end{equation}
If we solve the equations for $a(t),$ $b(t),$ and $c(t)$ we find
that the solutions repeat the initial conditions after a time $t_{R}=\pi/(\sqrt{2}\lambda)$
but at half that, $t_{H}=\pi/(2\sqrt{2}\lambda)$, the equations for
the operators are equivalent to detector equations for the interferometer
shown in Fig. \ref{figThreeWells}b. Initially we take a single particle
in each of the outer wells. The initial configuration repeats in $t_{R}$.
If a Fock state for the occupation of the three wells is $\left|m_{1},m_{2},m_{3}\right\rangle $,
then the wave function at time $t_{H}$ becomes
\begin{equation}
\left|\psi(t_{H})\right\rangle =-\frac{1}{\sqrt{8}}\left|200\right\rangle -\frac{1}{\sqrt{2}}\left|020\right\rangle -\frac{1}{\sqrt{8}}\left|002\right\rangle +\frac{1}{2}\left|101\right\rangle 
\end{equation}
Thus while the original state $\left|1,0,1\right\rangle $ is present,
the only other states are those with two particles in each well, in
a kind of HOM effect. 

\begin{figure}[h]
\centering \includegraphics[width=3.5in]{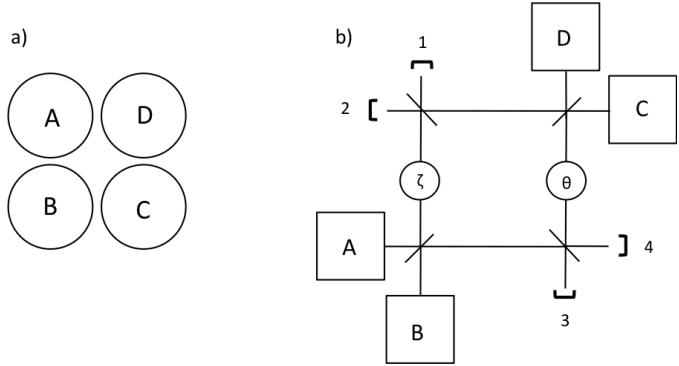}

\caption{(a) Four wells in a square with tunneling only along the sides. When
all four tunneling rates have the same value, $\lambda,$ the system,
at $t=\pi/4\lambda$, is equivalent to interferometer shown in (b)
with phase shifts $\zeta=\theta=0$. }

\label{figFourWells}
\end{figure}

We can place four wells on a square as shown in Fig. \ref{figFourWells}(a)
with particles initially only in wells 
A and C. Tunneling is between nearest-neighbor wells with the initial
state being \{A,B,C,D\} = \{1010\}. Possible states are \{2000, 0200,
0020, 0002, 1100, 1010, 1001, 0110, 0101, 0011\}. Now the repeat time
for the initial wave function is $\pi/2\lambda$. At half that, $t_{H}=\pi/4\lambda$,
the system is equivalent to the interferometer shown. With a single
particle in each of sites A and C, the wave function becomes 
\begin{equation}
\left|\psi(t_{H})\right\rangle =-\frac{1}{\sqrt{8}}\left(\left|2000\right\rangle +\left|0200\right\rangle +\left|0020\right\rangle +\left|0002\right\rangle \right)+\frac{1}{2}\left(\left|1010\right\rangle -\left|0101\right\rangle \right)
\end{equation}
 So in this case the interference removes occupation from these: \{1100\},
\{1001\}, \{0110\}, and \{0011\}, i.e., states with nearest neighbors
both occupied. 

\begin{figure}[h]
\centering \includegraphics[width=1.5in]{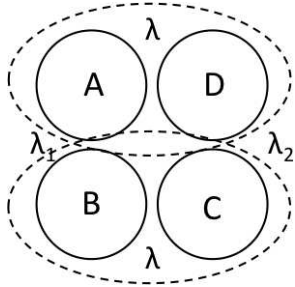}

\caption{Four wells with Alice's having counters A and D ; Bob has B and C.
The tunneling rate between A and D is $\lambda$ as is that between
B and C. The tunneling rate between A and B is $\lambda_{1}$ and
that between C and D is $\lambda_{2}.$ }

\label{figFourWellBell}
\end{figure}

However we can do a more interesting experiment with this form of
interferometer, namely look at a violation of the Bell theorem. An
interferometer method of testing the Bell theorem with arbitrary numbers
of bosons was proposed in Ref. \cite{LMBell}, with Alice and Bob
varying phase angles $\zeta$ and $\theta$ in a device like that
in Fig. \ref{figFourWells}(b). Here we consider the four-well system
with just a single particle in each of sites A and C. However, in
order to vary the phase angles in the four-well device, we have to
allow for \emph{variable} tunneling rates. Fig. \ref{figFourWellBell}
shows how we proceed and defines the variable tunneling rates $\lambda_{1}$
and $\lambda_{2}$. 
After a suitable time, $t=\pi/4\lambda$ we measure the parity product
$\left\langle P_{\alpha}P_{\beta}\right\rangle $ with particles in
A or C counting positive and those in B or D counting negative so
Alice's parity is $P_{\alpha}=(-1)^{n_{B}}$ and Bob's is $P_{\beta}=(-1)^{n_{D}}$.
We adjust $\lambda_{1}$ and $\lambda_{2}$ to form situations $P_{\alpha}^{\prime}$
and $P_{\beta}^{\prime}$ and maximize the usual quantity in the Bell-Clauser-Horne-Shimony-Holt
(BCHSH) form \cite{BCHSH} of the Bell theorem: $Q=\left\langle P_{\alpha}P_{\beta}\right\rangle +\left\langle P_{\alpha}^{\prime}P_{\beta}\right\rangle +\left\langle P_{\alpha}P_{\beta}^{\prime}\right\rangle -\left\langle P_{\alpha}^{\prime}P_{\beta}^{\prime}\right\rangle $.
Changing the tunneling rate is equivalent to a change of phase in
an arm of the interferometer. The tunneling rate $\lambda_{1}$ is
Alice's setting and $\lambda_{2}$ is Bob's. Denote a parity average
by $E(\frac{\lambda_{1}}{\lambda},\frac{\lambda_{2}}{\lambda})$.
We maximized $E(1,1)+E(1+\xi,1)+E(1,1-\xi)-E(1+\xi,1-\xi)$ and found
the maximum was 2.815 at $\xi=2.74.$ This result is very close to
the maximally possible violation of $2\sqrt{2}$ \cite{BellMax}.

\section{Conclusion}

We have shown how a simple set of potential-well traps can be used
to demonstrate the HOM effect and to build interferometers where the
beam splitter is provided by tunneling between wells. This suggested
method emphasizes the possibility of doing generalized HOM experiments
with \emph{multiple} particles by using condensates. The actual implementation
of such interference effects involves some requirements that could
be difficult: initially counting particle numbers in the wells when
used as sources, turning on tunneling for a specified time, and then
counting the particle numbers in the wells acting as detectors. Ref.
\cite{Kauf} has shown hat these difficulties can be overcome in the
$N=2$ case. When
interactions among the particles are considered we find the HOM effects
change considerably, showing the effects of possible ``fermionization'',
cooperative pair tunneling, and self-trapping that have been seen
in previous double-well analyses. We find that rearranging the wells
in various configurations can allow their use as a wide range of interferometers.

After completing our work we learned of related calculations by B.
Gertjerenken and P. G. Kevrekidis \cite{Panos}. We thank them for
sharing a preprint with us. We thank Dr. Asaad Sakhel for a critical
reading of the manuscript.

\section*{Appendix. Self-trapping}

We referred to self-trapping as a possible cause of the results shown
in Fig. 6. We would like to examine that claim a bit in this appendix.
Ref. \cite{SelfTrapB} discusses the dynamics of a Bose-Einstein condensate
in a double-well potential and the ``self trapping'' effect, which
occurs when the interactions between particles change the characteristics
of the tunneling between the wells. Within a mean-field approximation,
free tunneling between wells is reduced for interactions above a certain
critical value, causing an initial imbalance between numbers of particles
in the wells to be locked in with only smaller amplitude oscillations
away from that. In this approximation, all particles occupy the same
state given by a Gross-Pitaevskii equation 
\begin{equation}
\psi(\mathbf{r})=k_{1}u_{1}(\mathbf{r})+k_{2}u_{1}(\mathbf{r})
\end{equation}
where $u_{i}(\mathbf{r})$ represents the groundstate wave function
in well $i=1,2.$ The mean-field equations are expressions for the
amplitudes $k_{i.}$. When a more exact analysis is used (a full N-body
wave function depending on more than two parameters, but still a two-mode
approximation), quantum fluctuations alter the mean-field results
by time modulations and system revivals. 

If we derive equations of motion for the annhilation operators in
the Hamiltonian of Eq. (\ref{eq:H}), we find 
\begin{eqnarray}
i\frac{da}{dt} & = & -b+\gamma a^{\dagger}aa\nonumber \\
i\frac{db}{dt} & = & -a+\gamma b^{\dagger}bb
\end{eqnarray}
where we have measured energy from $E_{0}=0$ and time in units of
$\hbar/\lambda$ as before. If we replace $a^{\dagger}a$ by $N_{A}(t)$
and $b^{\dagger}b$ by $N_{B}(t)$ these are just the mean-field equations
for the $k_{i}$ found in Ref. \cite{SelfTrapB}, which, if one has
initial conditions $N_{A}(0)=N$, have the solution \cite{SelfTrapB}
\begin{equation}
N_{A}(t)=\frac{N}{2}\left[1+\mathrm{cn}(2t|N^{2}\gamma^{2}/16)\right]\label{eq:CN}
\end{equation}
where $\mathrm{cn}(u|v)$ is a Jacobi elliptic function. An equivalent
set of equations suitable for numerical solution for \emph{arbitrary}
initial conditions is given in Ref. \cite{SelfTrapA}. We can compare
our own exact results of Eqs. (\ref{eq:cnEqs}) with these mean-field
results. 

\begin{figure}[h]
\centering\includegraphics[width=2.5in]{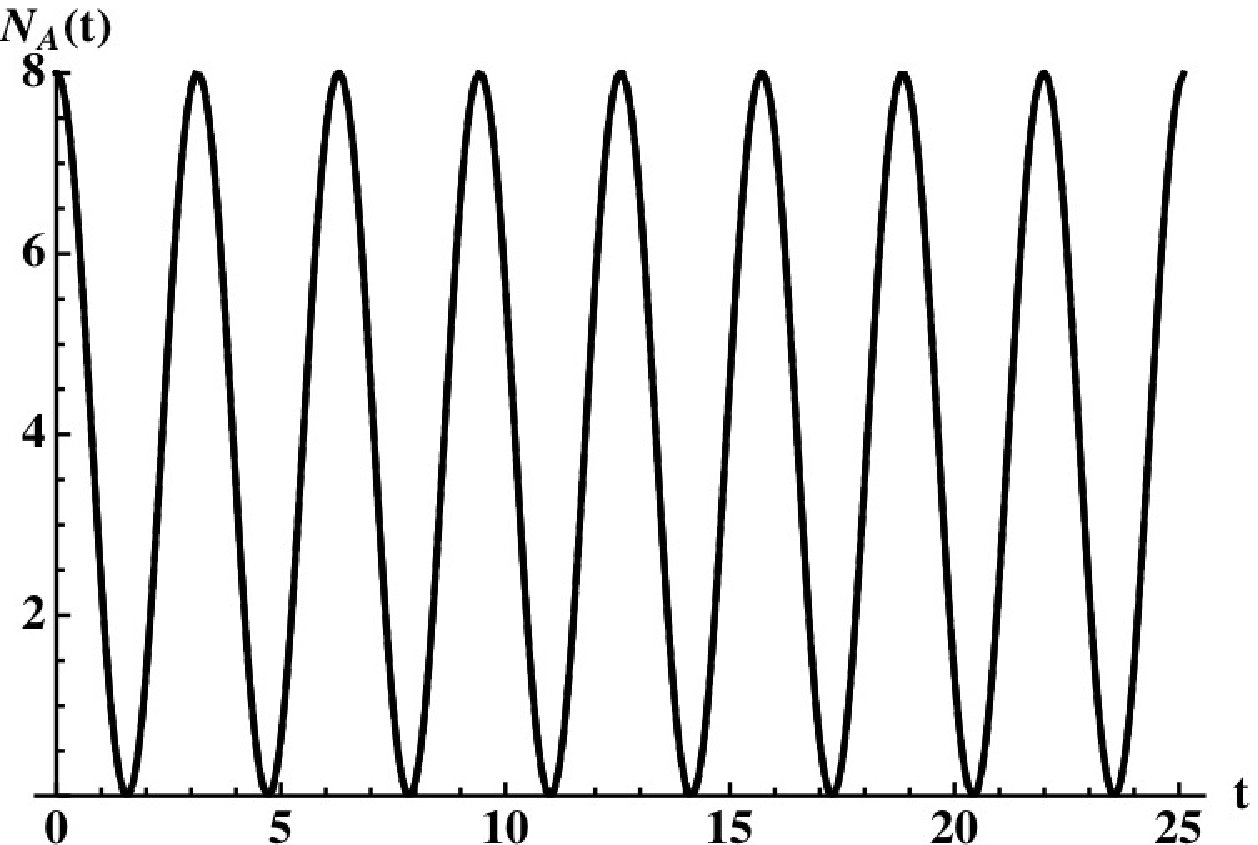}$\quad\quad$\includegraphics[width=2.5in]{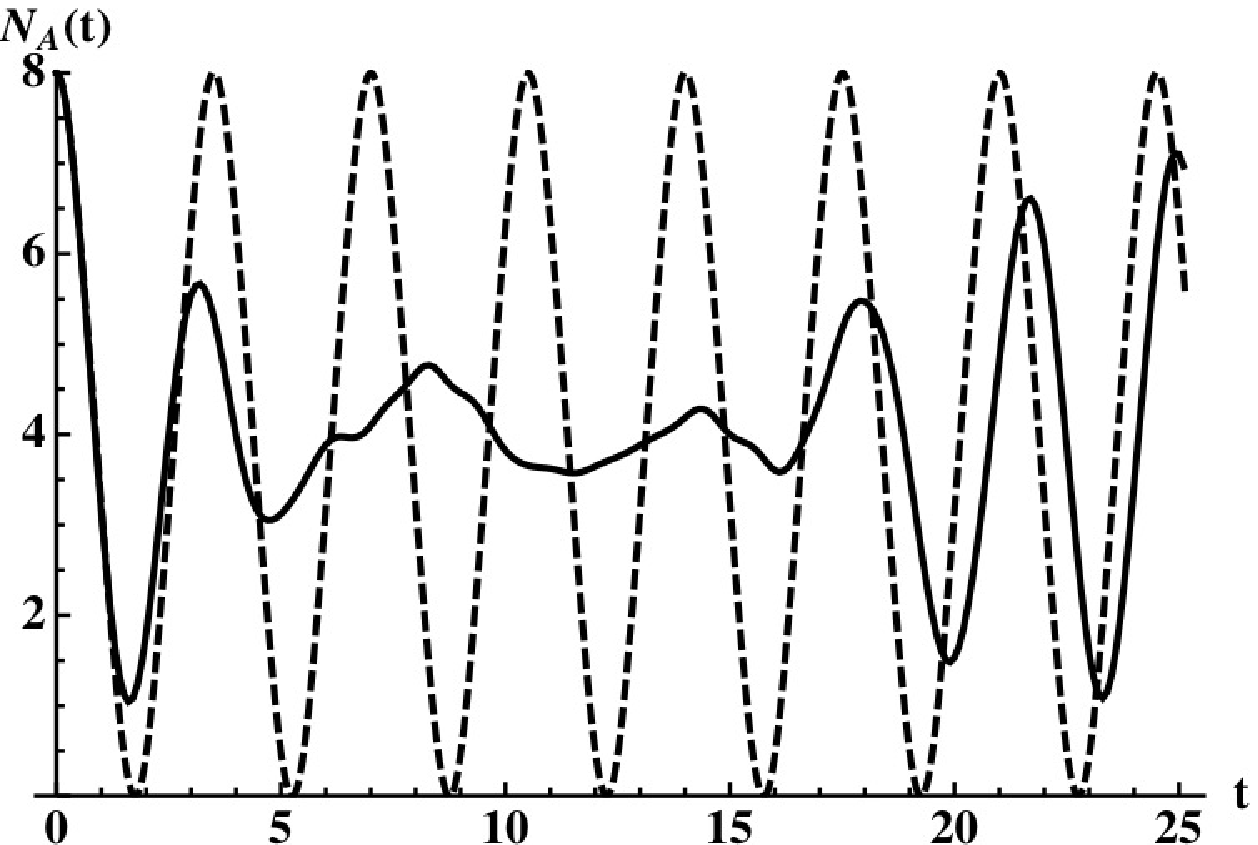}$\quad\quad$

\centering\includegraphics[width=2.5in]{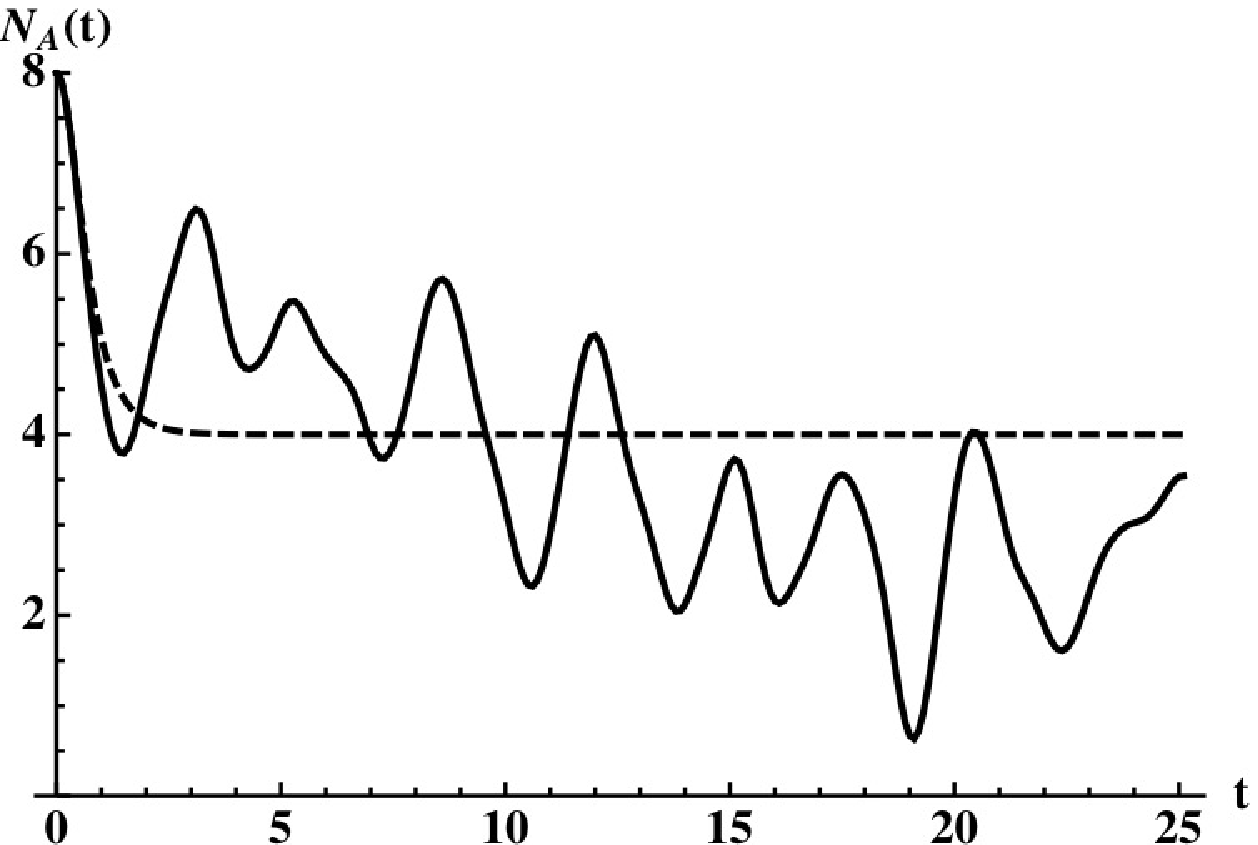}$\quad\quad$\includegraphics[width=2.5in]{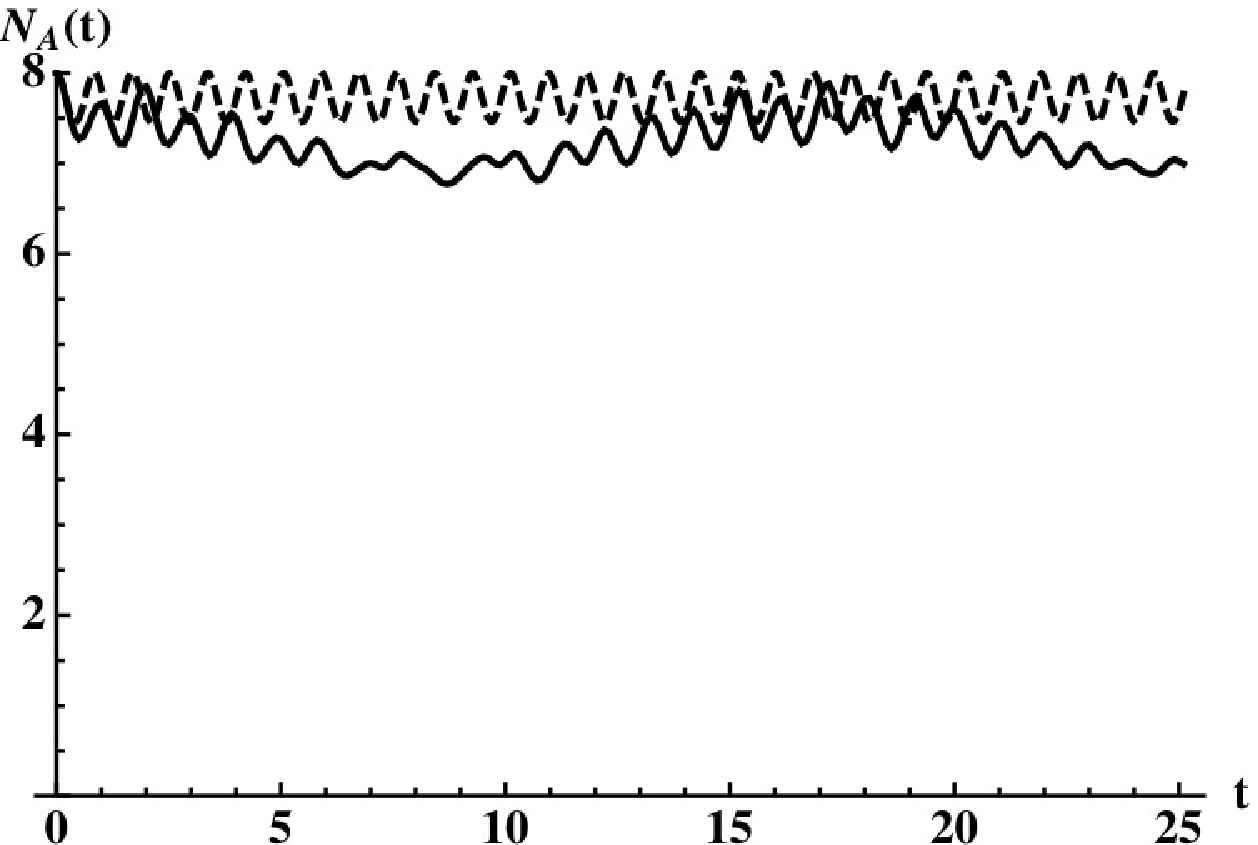}$\quad\quad$

\caption{Time dependence of $N_{A}(t)$ in a double-well potential with $N=8$
with initial condition $N_{A}(0)=N$. Solid line, exact result; dashed
line, mean-field result of Eq. (\ref{eq:CN}). (Upper Left) No interaction
$\gamma=0.$ The exact and mean-field results coincide. (Upper Right)
$\gamma=0.3$, below the self-trapping critical interaction value.
(Lower Left) $\gamma=0.5$ at the critical value. (Lower Right) $\gamma=1.0$,
above the critical value. }

\label{fig:SelfTrap}
\end{figure}

Note that in Sec. \ref{sec:HOM-calculations} our initial condition
was always $N_{A}(0)=N_{B}(0).$ In this case one finds in both exact
and mean-field cases that $N_{A}(t)=N/2$ because state $\{n,N-n\}$
is always paired in a superposition with state $\{N-n,n\}.$ (Note
the symmetry of the probabilities in Fig. \ref{fig:GammaN8}.) However,
self-trapping shows up dramatically when all the particles are initially
in one well, $N_{A}(0)=N$. The critical interaction for fixed $N$
is 
\begin{equation}
\gamma_{c}=\frac{4}{N}
\end{equation}
In Fig. \ref{fig:SelfTrap} we consider a small value $N=8$, although
our equations are valid for any $N$ value. The critical value then
is $\gamma_{c}=0.5.$ Our plots show the time dependence of $N_{A}(t).$
For $\gamma=0$ the mean-field and the exact value coincide with a
sinusoidal oscillation with full amplitude $0<N_{A}(t)<N$. At $\gamma=0.3<\gamma_{c}$
the mean-field continues to show full oscillation but an additional
fluctuation is introduced in the exact result. At critical value $\gamma=0.5$
the mean-field expression reaches self-trapping with $N_{A}(t)=N/2$;
the exact result fluctuates about this. (This result favoring equal
populations in both wells might be relevant in our own case of Fig.
\ref{fig:GammaN8}.) In the case of $\gamma>\gamma_{c}$ the mean-field
result maintains a value of $N_{A}(t)\ge N/2$ with oscillation amplitude
decreasing with increasing $\gamma$; the exact quantum value seems
to have an additional longer period oscillation frequency superposed,
similar to the two-frequency oscillations discussed in Ref. \cite{fermization}.
If $N_{A}(0)<N$ the amplitude of the oscillation below criticality
decreases correspondingly around $N/2.$

\begin{figure}[h]
\centering \includegraphics[width=3in]{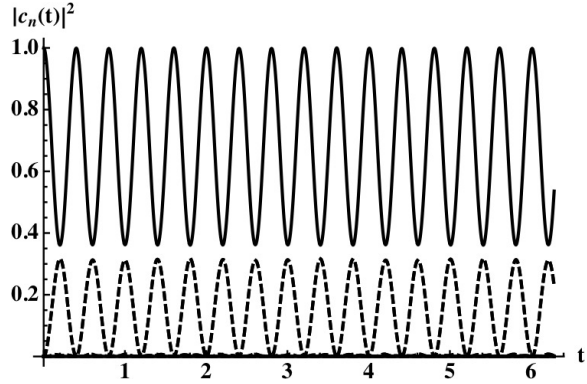}

\caption{Time dependence of configurations probabilities $\left|c_{n}(t)\right|^{2}$
in a double-well potential with $N=8$ and initial condition $N_{A}(0)=4$
with inteaction $\gamma=10$. The solid line gives $\left|c_{4}(t)\right|^{2}$
for initial configuration $\{4,4\}$, while the dashed line is for
$\left|c_{3}(t)\right|^{2}=\left|c_{5}(t)\right|^{2}$ for states
$\{3,5\}$ and $\{5,3\}$ respectively. All other states have very
small probabilities. The $\{4,4\}$ state is trapped at a relatively
large values. }

\label{figSelfTrapN4}
\end{figure}

None of these results exactly mirrors what we found in Fig. \ref{fig:GammaN8},
where the configuration \{4,4\} became favored. In that case, the
initial condition was $N_{A}(0)=4$, for which the symmetry of the
probabilites leads to $N_{A}(t)=4$, i.e., constant in value, both
exactly and in mean-field approximation. However, if we look at the
details of the probabilities $p_{n}(t)=\left|c_{n}(t)\right|^{2}$
for individual configurations $\{n,N-n\},$ then we find that the
initial configuration itself can become ``trapped'' at a large value,
as seen in Fig. \ref{figSelfTrapN4} for a large interaction parameter.
This effect is analogous to the ``fermionization'' we saw for just
two particles in Fig. \ref{fig:Fermionization}. This effect
is due simply to fact that the initial condition enters with approximately
unit amplitude and that any other states are mixed in only in higher
order in $1/\gamma$. 

This is an effect that cannot be seen in the mean-field calculations;
we plan to return to this subject in a future publication.\clearpage{}

\end{document}